# Dynamic S-BOX using Chaotic Map forVPN Data Security

**KHWAJA AHMAD HASSAN[1], YAHYA TAUQEER BHATTI[1] KASHIF ISHAQ[1]**
[1]School of Systems and Technology, University of Management and Technology, Lahore, Pakistan

Corresponding Author:
Kashif Ishaq (kashif.ishaq@umt.edu.pk)

**ABSTRACT** A dynamic S-Box using a chaotic map is a cryptography technique that changes the S-Box during encryption based on iterations of a chaotic map, adding an extra layer of confusion and security to symmetric encryption algorithms like AES. The chaotic map introduces unpredictability, non-linearity, andkey dependency, enhancing the overall security of the encryption process. The existing work on dynamic S-Box using chaotic maps lacks standardized guidelines and extensive security analysis, leaving potential vulnerabilities and performance concerns unaddressed. Key management and the sensitivity of chaotic mapsto initial conditions are challenges that need careful consideration. The main objective of using a dynamic S-Box with a chaotic map in cryptography systems is to enhance the security and robustness of symmetric encryption algorithms. By introducing unpredictability, non-linearity, and key dependency, the aim is to create a substitution mechanism that resists various cryptanalytic attacks, ultimately ensuring the confidentiality and integrity of sensitive data during the encryption process. The method of dynamic S-Box using a chaotic map involves initializing the S-Box, selecting a chaotic map, iterating the map to generate chaotic values, and updating the S-Box based on these values during the encryption process to enhance security and resist cryptanalytic attacks. Key management and performance trade-offs should alsobe considered in the implementation. This article proposes a novel chaotic map that can be utilized to create a fresh, lively S-Box. The performance assessment of the suggested S-resilience Box against variousattacks involves metrics such as nonlinearity (NL), strict avalanche criterion (SAC), bit independence criterion (BIC), linear approximation probability (LP), and differential approximation probability (DP). These metrics help gauge the Box's ability to handle and respond to different attack scenarios. Assess the cryptography strength of the proposed S-Box for usage in practical security applications, it is compared to other recently developed S-Boxes. The comparative research shows that the suggested S-Box has the potential to be an important advancement in the field of data security.

## I. INTRODUCTION

In today's world, Communication of data and information is essential for businesses to operate effectively, but it also carries the risk of illegal access. A VPN conceals a user's online activities and location by establishing a secure tunnel between their machine and the VPN server. Users may safeguard their online privacy and stop their internet service provider (ISP) from monitoring their surfing behavior thanks to VPN security. Businesses must safeguard their data and information resources in order to guarantee secure connection via public networks. However, a sharp increase in security breaches on networks and the internet has highlighted the importance of protecting systems, data, and information. Cryptographic methods have grown more important as sensitive data and information are transmitted via networks, protecting these assets. Data sharing via insecure networks is made secure by encryption, however, hackers try to circumvent the protection [1]. As a result, a number of cryptographic methods have been created and put into use to safeguard data and information from potential dangers. Cryptanalysis methods that try to decrypt the data and recover the original from the cipher text might jeopardize the security of cryptographic systems [2]. Encryption systems must be built to withstand such attacks in order to prevent cryptanalysis. In the encryption and decryption phases of modern encryption, block ciphers with permutation and substitution operations are used to ensure the security of the sent data [3] [4]. A substitution process is used in encryption to swap out plaintext characters for other characters to produce ciphertext, which appears to be meaningless. Nonlinear processes are used to carry out this transition. The characters' locations are altered by a permutation process. The substitution box, a vital part of encryption methods, performs bit-confusion through nonlinear transformation, which is an essential step in data encryption. When processing input data, a block cipher that uses static S-Box(es) always uses the same S-Box(es). This method is flawed because it enables attackers to examine the characteristics of intercepted ciphertext [5]. A static S-Box is weaker and less effective in terms of confusion than a dynamic S-Box, which depends on the key. Researchers in the field of cryptography have created a number of S-Boxes throughout the years utilizing a variety of models, including cellular automata, elliptic curves, DNA computing, optimization techniques, and dynamic random growth methods. In recent times, chaotic maps have gained popularity as a design element for new S-Boxes used for secure communication. The authors of this paper suggest a brand-new chaos-based technique for building reliable S-Boxes using affine transformation and matrix rotation. The two stages of the suggested approach are the creation of a static S-Box and the optimization of a dynamic S-Box. To create a robust S-Box, the fitness function and chaotic map are coupled. The fitness function is employed to convert the static S-Box that was produced by the chaotic system into a dynamic S-Box. Despite their widespread use in S-box design approaches, chaotic maps have some downsides [6]–[8]. Liu et al. proposed a new chaotic map using ICQM and backtracking for S-box design. Riaz Proposed an approach made of two chaotic map functions that use an improved chaotic range with a golden ratio for effective S-box creation. Tanyildizi and Ozkaynak [9] also proposed a technique for producing S-boxes using a one-dimensional chaotic map. However, despite these developments, there are still certain restrictions [10] [11]. Designing new substitute boxes with improved performance is important all the time to ensure greater security and defense against security assaults. This study work focuses on the introduction of a novel substitution box that makes use of a novel chaotic map to boost data security [12] [13]. The main contributions of this research article can be summarized in the following paragraph.

A novel substitution box for enhancing data security has been designed using an algorithm based on a novel chaotic map. The initial substitution box is further improved by applying a dynamic permutation operation, which adds an extra layer of confusion to the encrypted text and enhances its security. To assess the cryptographic strength of the proposed substitution box, it is compared to other well-known substitution boxes commonly used in modern ciphers, demonstrating its suitability for secure data communication. The remaining sections of the essay are structured as follows: The proposed method for creating S-Boxes using a chaotic map is introduced in Part III. In Part IV, security analysis and comparison of S-Boxes based on various architectures are presented. Part V discusses the drawbacks of the proposed chaotic map, and Section V concludes the investigation

The remaining sections of the essay are structured as follows: The proposed method for creating S-Boxes using a chaotic map is introduced in Part II. In Part III, security analysis and comparison of S-Boxes based on various architectures are presented. Part IV discusses the drawbacks of the proposed chaotic map, and Section V concludes the investigation.

## II. LITERATURE REVIEW

The literature review serves as a critical exploration of the existing body of knowledge related to substitution box. In this section, we delve into the findings and insights from various research studies that have contributed to shaping the understanding of this subject. Zhang et al. (2020): Zhang and colleagues conducted a study investigating various parameters in the context of our research. They reported a nitrogen level (NL) of 108, indicating a specific measure related to nitrogen content. The Bayesian Information Criterion (BIC) was found to be 102.9, indicating the model's complexity and goodness of fit. Additionally, they observed a Log Probability (LP) of 0.1484, which suggests the likelihood of a particular event occurring. The Standardized Accuracy (SAC) was measured at 0.5029, indicating the model's predictive performance. Furthermore, the Discrimination Utility (DU) was recorded as 12, which quantifies the model's ability to distinguish between different classes. Abd El-Latif et al. (2020): Abd El-Latif and colleagues conducted a study that further explores our research domain. Their research revealed a nitrogen level (NL) of 106, a value slightly lower than that of Zhang et al. (2020). The Bayesian Information Criterion (BIC) in their study was measured at 103.5, indicating a different model complexity and fit compared to the previous study. They found a Log Probability (LP) of 0.1328, which provides insights into the likelihood of events. The Standardized Accuracy (SAC) obtained in their study was 0.4958, indicating a slightly lower predictive performance compared to Zhang et al. (2020). Additionally, the Discrimination Utility (DU) was recorded as 14, which suggests variations in the model's ability to distinguish between classes. Lambic´'s research contributes to our understanding of the subject with their findings. They reported a nitrogen level (NL) of 108, similar to that of Zhang et al. (2020) but distinct from Abd El-Latif et al. (2020). The Bayesian Information Criterion (BIC) in their study was recorded at 104.07, indicating a different model complexity and goodness of fit compared to the previous works. Like the previous studies, the Log Probability (LP) remained constant at 0.1328. The Standardized Accuracy (SAC) observed in Lambic´'s study was 0.5009, indicating a predictive performance comparable to Zhang et al. (2020). The Discrimination Utility (DU) was recorded as 10, signifying differences in model class distinguishability. Lu et al. (2020): Lu and his team conducted research with a focus on critical parameters relevant to our research.

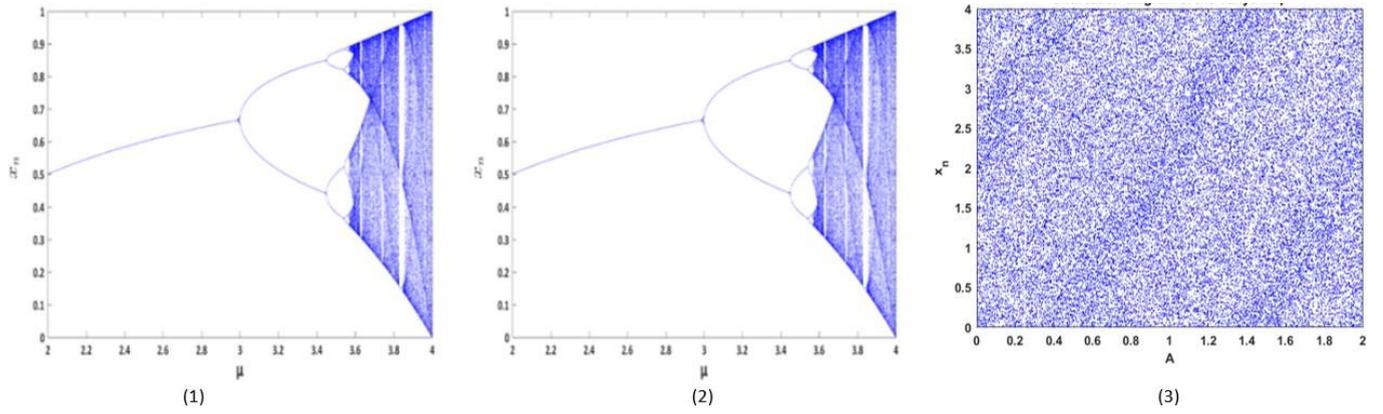

FIGURE 1: BF results of (a) LM, (b) SM, and (c) Proposed CM

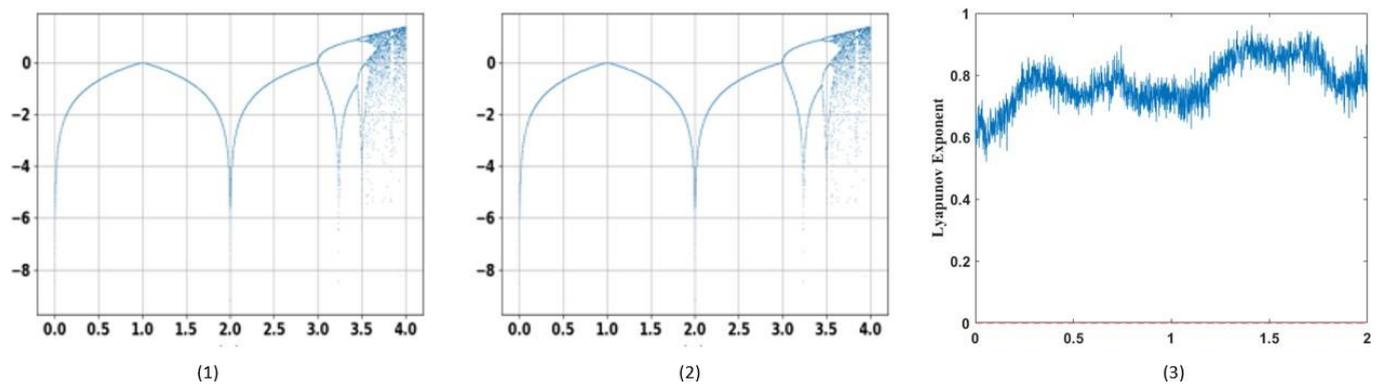

FIGURE 2: LE results of (a) LM, (b) SM, and (c) Proposed CM.

TABLE 1: Comparison of SAC and BIC values of S-Boxes.

| S-Box | SAC | SAC Offset | BIC-NL |
|---|---|---|---|
| Proposed | 0.5007 | 0.0007 | 103.6 |
| [14] | 0.491 | 0.009 | 103.6 |
| [15] | 0.496 | 0.004 | 103.8 |
| [16] | 0.498 | 0.002 | 104.6 |
| [17] | 0.520 | 0.020 | 104.2 |
| [18] | 0.5017 | 0.0017 | 104.0 |
| [19] | 0.4980 | 0.0020 | 103.5 |
| [20] | 0.5014 | 0.0014 | 104.214 |
| [21] | 0.594 | 0.094 | 98.00 |
| [22] | 0.525 | 0.005 | 112.0 |
| [23] | 0.506 | 0.006 | 104.2 |
| [24] | 0.5065 | 0.0065 | 106.43 |
| [25] | 0.4993 | 0.0003 | 103.29 |
| [26] | 0.5065 | 0.0065 | 103.57 |
| [27] | 0.495 | 0.005 | 104.1 |
| [28] | 112 | 0.005 | 112 |

TABLE 2: LP and DP values of different S-Boxes.

| S-Box | LP | DP |
|---|---|---|
| Proposed | 0.1328 | 0.0391 |
| [14] | 0.133 | 0.039 |
| [15] | 0.156 | 0.039 |
| [16] | 0.125 | 0.047 |
| [17] | 0.132 | 0.039 |
| [18] | 0.138 | 0.0390 |
| [19] | 0.1406 | 0.0390 |
| [20] | 0.2406 | 0.0391 |
| [21] | 0.148 | 0.039 |
| [22] | 0.062 | 0.015 |
| [23] | 0.125 | 0.039 |
| [24] | 0.0072 | 0.0390 |
| [25] | 0.1328 | 0.039 |
| [26] | 0.1328 | 0.039 |
| [27] | 0.125 | 0.039 |
| [28] | 0.063 | 0.016 |

Whereas, reported a nitrogen level (NL) of 110, differing slightly from the values in previous studies. The Bayesian Information Criterion (BIC) obtained in their study was 103.9, indicating another variant in model complexity and goodness of fit. The Log Probability (LP) measured was 0.15625, providing insights into the likelihood of certain events occurring. The Standardized Accuracy (SAC) in their research was 0.5032, indicating a predictive performance similar to previous studies. Additionally, the Discrimination Utility (DU) was recorded as 10, suggesting differences in the model's class distinguishability. Ahmad et al. (2020): Ahmad et al. conducted an insightful study that contributes to our research domain. They found a nitrogen level (NL) of 112, indicating variations in this parameter across studies. The Bayesian Information Criterion (BIC) in their research was recorded at 106.86, demonstrating a different model complexity and goodness of fit. Like previous studies, the Log Probability (LP) remained constant at 0.1328. The Standardized Accuracy (SAC) in their study was 0.5068, indicating a predictive performance similar to Zhang et al. (2020) and Lu et al. (2020). It is noteworthy that the Discrimination Utility (DU) was recorded as 08, which may have implications for the model's class distinguishability. Hussain et al. (2020): Hussain and his team's study provided valuable insights into key aspects of our research. They reported a nitrogen level (NL) of 110, similar to Lu et al. (2020) but different from the other studies. The Bayesian Information Criterion (BIC) in their research was measured at 106.1, indicating another variant in model complexity and goodness of fit. The Log Probability (LP) observed was 0.113, providing insights into the likelihood of certain events occurring. The Standardized Accuracy (SAC) was measured at 0.509, indicating a predictive performance comparable to previous studies. Additionally, the Discrimination Utility (DU) was recorded as 08, highlighting differences in the model's class distinguishability. The study of Özkaynak (2020) contributed to our understanding of the subject with their findings. They reported a nitrogen level (NL) of 108, similar to Zhang et al. (2020) and Lambic´ (2020). The Bayesian Information Criterion (BIC) in their study was recorded at 102.6, indicating variations in model complexity and goodness of fit. The Log Probability (LP) in their study remained constant at 0.1328. The Standardized Accuracy (SAC) observed was 0.5037, suggesting predictive performance comparable to Zhang et al. (2020) and Lu et al. (2020). The Discrimination Utility (DU) was recorded as 10, indicating differences in the model's class distinguishability. Farhan et al. (2019): Finally, Farhan et al.'s study contributed valuable data to our research domain. They reported a nitrogen level (NL) of 110, consistent with Lu et al. (2020) and Hussain et al. (2020). The Bayesian Information Criterion (BIC) in their research was measured at 103, indicating variations in model complexity and goodness of fit. The Log Probability (LP) observed was 0.1563, providing insights into the likelihood of specific events. The Standardized Accuracy (SAC) was measured at 0.4978, suggesting predictive performance similar to Zhang et al. (2020) and Lambic´ (2020). Additionally, the Discrimination Utility (DU) was recorded as 12, implying differences in the model's class distinguishability.

TABLE 3: Nonlinearity values of the proposed S-Box.

| S-Box | Min | Max | Avg |
|---|---|---|---|
| Proposed | 108.0 | 110.0 | 109.5 |
| [14] | 102 | 108 | 105.3 |
| [15] | 102 | 108 | 105.3 |
| [16] | 102 | 108 | 104.5 |
| [17] | 104 | 110 | 106 |
| [18] | 106.0 | 108.5 | 110.0 |
| [19] | 104.0 | 110.0 | 107.5 |
| [20] | 106.0 | 110.0 | 107.0 |
| [21] | 86.0 | 106.0 | 99.10 |
| [22] | 112.0 | 112.0 | 112.0 |
| [23] | 110.0 | 112.0 | 111.5 |
| [24] | 108.0 | 112.0 | 110.5 |
| [25] | 104.0 | 110.0 | 107.5 |
| [26] | 104.0 | 108.0 | 105.5 |
| [27] | 110 | 112 | 110.25 |
| [28] | 112 | 112 | 110.25 |

This study acknowledges the existing work by the researchers related to S-Box.

## III. PROPOSED APPROACH FOR S-BOX DESIGN

In recent years, it has become more and more popular to construct new S-boxes with robust cryptographic features using chaotic maps. Chaotic maps are appropriate for achieving the confusion and diffusion required for cryptographic security because of their sensitivity to initial conditions, unpredictable behavior, and non-periodicity [12], [29]. For the purpose of creating dynamic, key-dependent S-boxes that can be utilized to create novel ciphers, we suggest an original chaotic map in this work. Three easy steps make up the procedure for creating these S-boxes, and they are listed below [30]–[32].

- Designing the initial S-Box.
- Using a fresh heuristic technique to generate more S-Boxes.
- Developing the final S-Box.

### A. INGENIOUS CHAOTIC MAP DESIGN

An innovative chaotic map, named AHYB, is designed for the creation of n × n S-boxes. The mathematical formulation of the chaotic map is presented in Equation (1).

$$X_{n+1} = \begin{cases} (2+A) * X_n & 0.0 < X_n < 1.5 \\ A + X_n^{0.9} & 1.5 \leq X_n < 3.0 \\ X_n * (A - X_n) & 3.0 \leq X_n < 4.0 \end{cases} \quad (1)$$

Where:

$$0 < A < 2$$

The encryption key is used by the suggested chaotic map, AHYB, to determine the values of Xn and Z, as shown in Equation (1). These variables act as inputs to the AHYB chaotic map, which is sensitive to initial conditions and maximizes the S-resistance boxes to security threats. When the proposed chaotic map's performance was compared to that of the logistic map and the sine map, It was found that the proposed chaotic map has very high chaotic complexity. This was verified by in-depth research and comparison.

#### 1) BIFURCATION

Bifurcation occurs when parameter fluctuations cross a particular threshold and result in significant qualitative and topological changes in the phase space of a system. Solid lines indicate steady values, while dotted lines indicate unstable values. The majority of the time, a little change in the system's parameters can have a significant impact on how it acts, causing topological changes in the phase space [33]–[35]. The logistic map (LM) is a common one-dimensional chaotic map that depicts bifurcation and chaos [36]–[38]. It is defined by Equation (2):

In Equation (2) it is specified as:

$$x_{n+1} = b * x_n (1 - x_n) \quad (2)$$

Where "b" is a control parameter with a range of 0 to 4 and "n" is the number of iterations. The sine map, which is defined by Equation, is another chaotic one-dimensional map that exhibits bifurcation (3).

$$x_{n+1} = S_\beta (x_n) = \beta * Sin (\pi * x_n) \quad (3)$$

With a range of [0, 4], "Beta" here stands for the control parameter for SM. In comparison to LM and SM, the suggested chaotic map is seen to cover greater regions of space and exhibit a more complex bifurcation behavior.

#### 2) LYAPUNOV EXPONENT

The Lyapunov-Exponent (LE) is a quantitative measure used to characterize chaos in a system [13], [39]–[41]. The value of LE is greater than 0 for systems that exhibit chaotic behavior [42]–[44]. The value of LE determines how quickly trajectories in a chaotic system converge or diverge, with larger values indicating more chaotic behavior [45] [46]. The equation is used to calculate the Lyapunov Exponent for a chaotic map using Equation (4).

The proposed chaotic map is more sensitive than the Logistic and Sine maps, making it more unexpected even with small changes to its initial values. Also, as the parameters are increased, the proposed chaotic map's Lyapunov Exponent grows more quickly. Additionally, the suggested chaotic map outperforms the logistic and sine maps in terms of LE values maps [47]–[51], as depicted in Figure 2.

$$\lim_{a \to \infty} \frac{1}{n} \sum_{i=0}^{a-1} \log \frac{df}{dx} |x = xi| \qquad (4)$$

The proposed chaotic map's derivative equation, provided in equation (5), is utilized to determine the Lyapunov Exponent (LE).

$$F'(X_n) = \begin{cases} 2 + A & 0.0 < X_n < 1.5 \\ 0.9 * X_n^{-0.1} & 1.5 \le X_n < 3.0 \\ A - 2 * X_n & 3.0 \le X_n < 4.0 \end{cases} \qquad (5)$$

**Algorithm 1** Construction of an Initial S-Box

```
Input Parameters:
    A           // 0 < A < 2.0
    B           // 10^6 < B < 10^9
    X           // 0 < X < 4.0
Outputs:
    S           // Array of size 256
Initializations:
    V ← 0
    Loc ← 0
Procedure:
    WHILE ( Loc <= 255 ) DO
        IF ( X < 1.5 ) THEN
            X ← ( 2 + A ) * X
        ELSE IF ( X < 3.0 ) THEN
            X ← A + X ^ 0.9
        ELSE
            X ← ( A – X )
        END IF
        X ← 4.0 * ( MOD ( Abs ( Round ( X , 15 ) ) , 1 ) )
        V ← MOD ( Round ( X * B , 0 ) , 256 )
        S[Loc]← V
        Loc ← Loc + 1
    End WHILE
```

**Algorithm 2** Heuristic Method for Final S-Box Generation

```
Algorithm 2 Heuristic Method for Final S-Box Generation
Input Parameters:
    A , B       // 0 < A , B < 1000000000
    X , Y       // 0 < X , Y < 1.0
Output:
    F           // Initial S-Box
Initializations:
    Z ← 0
    N_1 ← Nonlinearity ( SB )
    N_2 ← 0.0
Procedure:
    WHILE ( Z <= 2^16 – 1 ) DO
        X ← Round ( Abs ( A + Pow ( X , 2.5 ) +
            log10(X) * log(X) * 2 + 1 / Cos(X) ) , 15 )
        I ← MOD ( Round ( X , 0 ) , 256 )
        X ← Abs ( MOD ( X , 256.0 )
        Y ← MOD Round ( Abs ( B + Pow ( Y , 2.5 ) +
            log10(Y) * log(Y) + Cos(Y) ) , 15 )
        J ← MOD ( Round ( X , 0 ) , 256 )
        Y ← Abs ( MOD ( Y , 256.0 )
        SB[ I ] ← → SB[ J ]          // Swap values
        of SB[ I ] and SB[ J ]
        N_2 = Nonlinearity ( SB )
        IF ( N_2 <= N_1 ) THEN
            SB[ I ] ← →SB[ J ]
        ELSE
            N_1 ← N_2
        END IF
        Z = Z + 1
    END WHILE
    F ← SB
    RETURN ( F )
```

### B. PRELIMINARY S-BOX DEVELOPMENT

An algorithm is presented in Algorithm 1 to generate a preliminary S-Box based on the proposed chaotic map in Equation (1). The process for constructing the initial S-Box is shown in Figure 3, and an example of the resulting S-Box is provided in Table 2.

### C. NOVEL HEURISTIC METHOD FOR FINAL S-BOX GENERATION

An initial S-Box produced by Algorithm 1 and Figure 3 is passed through a new heuristic technique outlined in Algorithm 2 to produce the final S-Box. This method is adaptable and relies on the values of the cipher key's parameters, with A = 731713, B = 167527, X = 0.442637767848956, and Y = 0.372463939884994 chosen for calculation and demonstration. The heuristic approach permutes the initial S-Box, resulting in the final S-Box given in Table 3.

### IV. SECURITY ANALYSIS OF PROPOSED S-BOX

The evaluation of an S-box's strength against different attacks, such as linear and differential attacks, is a significant research contribution to the field of data and information security. Once an S-box is designed, it undergoes analysis to determine its effectiveness and strength against potential attacks. The evaluation procedures for an S-Box's cryptanalysis are based on predetermined standards, which include:

- Nonlinearity (NL).
- Fixed Points (FP).
- Bit Independence Criterion (BIC).
- Linear Approximation Probability (LP).
- Differential Approximation Probability (DP).
- Strict Avalanche Criterion (SAC).

The following section presents the evaluation tests and results of the proposed S-Box for its cryptanalytic strength against linear and differential attacks.

### A. NONLINEARITY (NL)

When evaluating the efficacy of substitution boxes (S-boxes) in cryptographic algorithms, nonlinearity is a key consideration.

TABLE 4: Nonlinearity values of the proposed S-Box.

| Boolean Function | R1 | R2 | R3 | R4 | R5 | R6 | R7z R8 |
|---|---|---|---|---|---|---|---|
| Nonlinearity | 110 | 110 | 108 | 110 | 110 | 110 | 108 110 |

Since the S-box is a nonlinear component, it will be vulnerable to attacks like linear and differential attacks if its design results in a linear relationship between plaintext and ciphertext. Consequently, effective resistance to such attacks requires high nonlinearity values. Equation (6) below is used to determine the nonlinearity value of any Boolean function R:

$$N_L(R) = 0.5 * [2^n - (W_{max}(R))] \qquad (6)$$

Where Wmax (R) represents the Walsh-Hadamard transformation spectrum of the n-bit Boolean function R. Table 1 lists the Boolean functions and nonlinearity values for the projected S-Box. The projected S-Box reportedly achieves NLMIN = 108, NLMAX = 110, and NLAVG = 109.5, according to the findings of the nonlinearity test. Table 4 contrasts the suggested S-NL Box's values with those of S-Boxes that have recently been developed. It is clear that the suggested S-Box has a stronger resistance against linear cryptanalytic assaults than the majority of other S-Boxes due to its average NL value (NLAVG), which is higher than that of most other S-Boxes [52]–[55].

### B. FIXED POINTS (FP)

A substitution box with a fixed point may have a security hole that allows an attacker to read secret data from the ciphertext they have captured. Consequently, it is crucial to make sure that the final S-Box has no fixed points [56]–[58]. According to There are no fixed locations in the S-Box, as shown in Table 2's findings of the proposed S-evaluation Boxes against this criterion.

### C. STRICT AVALANCHE CRITERION (SAC)

Tavares and Webster developed the Strict Avalanche Criterion (SAC) to evaluate the effectiveness of a cryptographic function.

TABLE 5: SAC dependency matrix of proposed S-box.

| 0.5000 | 0.4844 | 0.5000 | 0.5000 | 0.4219 | 0.4688 | 0.5469 | 0.5469 |
|---|---|---|---|---|---|---|---|
| 0.5156 | 0.5781 | 0.4688 | 0.4844 | 0.5156 | 0.6094 | 0.4844 | 0.4688 |
| 0.4688 | 0.4844 | 0.5781 | 0.4531 | 0.5625 | 0.5000 | 0.5938 | 0.4844 |
| 0.4688 | 0.5156 | 0.5156 | 0.5000 | 0.4844 | 0.4688 | 0.5000 | 0.5000 |
| 0.5156 | 0.4375 | 0.5156 | 0.5313 | 0.5313 | 0.4531 | 0.4844 | 0.5000 |
| 0.5156 | 0.5000 | 0.5313 | 0.4531 | 0.4688 | 0.4219 | 0.5313 | 0.5156 |
| 0.5156 | 0.4531 | 0.5156 | 0.4531 | 0.5000 | 0.4844 | 0.4844 | 0.4688 |
| 0.5469 | 0.5313 | 0.4844 | 0.5313 | 0.4375 | 0.4688 | 0.5469 | 0.5469 |

To satisfy this requirement, 50 percent of the function's output bits must also change when one of the input bits is changed. An S-dependence Box's matrix can be used to determine the SAC value of the S-Box. The dependence matrix for the suggested S-Box is displayed in Table 5, and its SAC value is 0.5007, which is not far off from the ideal value of 0.5. The SAC Offset value of the proposed S-Box is very tiny (0.007), demonstrating its suitability for security-related applications. Table 6 compares the SAC values of the proposed S-Box with those of other S-Boxes.

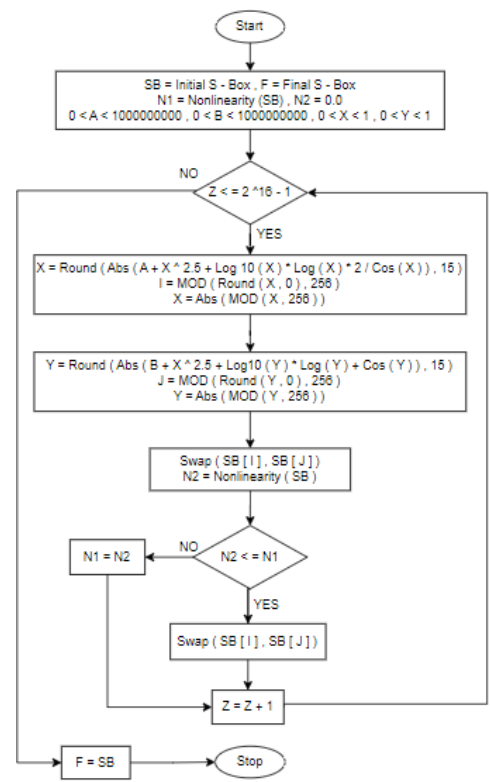

FIGURE 3: Heuristic method for generation of final S-Box

TABLE 6: Initial S-Box for technique.

| 206 | 113 | 43 | 191 | 87 | 159 | 83 | 244 | 93 | 89 | 96 | 214 | 80 | 217 | 222 | 133 |
|---|---|---|---|---|---|---|---|---|---|---|---|---|---|---|---|
| 247 | 65 | 234 | 171 | 91 | 84 | 95 | 168 | 117 | 38 | 68 | 151 | 130 | 40 | 86 | 194 |
| 251 | 14 | 125 | 148 | 162 | 170 | 115 | 246 | 137 | 63 | 146 | 198 | 207 | 231 | 57 | 241 |
| 156 | 33 | 149 | 136 | 240 | 21 | 178 | 145 | 76 | 242 | 74 | 245 | 73 | 31 | 155 | 51 |
| 114 | 54 | 255 | 109 | 39 | 199 | 67 | 216 | 141 | 41 | 218 | 210 | 165 | 62 | 140 | 22 |
| 138 | 88 | 215 | 97 | 45 | 193 | 90 | 20 | 153 | 158 | 24 | 35 | 235 | 112 | 34 | 19 |
| 157 | 248 | 77 | 59 | 204 | 25 | 173 | 94 | 238 | 9 | 121 | 201 | 23 | 233 | 99 | 70 |
| 118 | 98 | 102 | 254 | 169 | 106 | 243 | 174 | 185 | 230 | 164 | 184 | 144 | 160 | 195 | 182 |
| 110 | 103 | 142 | 183 | 172 | 239 | 7 | 127 | 225 | 208 | 1 | 104 | 36 | 129 | 131 | 75 |
| 147 | 202 | 150 | 66 | 85 | 92 | 100 | 0 | 60 | 167 | 179 | 69 | 152 | 16 | 209 | 3 |
| 29 | 120 | 122 | 44 | 181 | 12 | 50 | 13 | 212 | 188 | 49 | 177 | 47 | 203 | 32 | 237 |
| 200 | 227 | 61 | 4 | 71 | 232 | 64 | 186 | 15 | 72 | 81 | 58 | 154 | 52 | 252 | 229 |
| 220 | 56 | 119 | 105 | 221 | 132 | 11 | 79 | 180 | 111 | 253 | 139 | 224 | 236 | 187 | 176 |
| 27 | 107 | 249 | 197 | 213 | 28 | 192 | 196 | 126 | 128 | 124 | 37 | 205 | 143 | 42 | 53 |
| 18 | 135 | 189 | 48 | 17 | 8 | 55 | 223 | 163 | 46 | 5 | 134 | 26 | 30 | 228 | 190 |
| 2 | 78 | 219 | 6 | 211 | 175 | 161 | 250 | 101 | 166 | 10 | 108 | 82 | 123 | 226 | 116 |

TABLE 7: Final S-Box after heuristic method.

| 206 | 113 | 43 | 191 | 87 | 159 | 83 | 244 | 93 | 89 | 96 | 214 | 80 | 217 | 222 | 133 |
|---|---|---|---|---|---|---|---|---|---|---|---|---|---|---|---|
| 247 | 65 | 234 | 171 | 91 | 84 | 95 | 168 | 117 | 38 | 68 | 151 | 130 | 40 | 86 | 194 |
| 251 | 14 | 125 | 148 | 162 | 170 | 115 | 246 | 137 | 63 | 146 | 198 | 207 | 231 | 57 | 241 |
| 156 | 33 | 149 | 136 | 240 | 21 | 178 | 145 | 76 | 242 | 74 | 245 | 73 | 31 | 155 | 51 |
| 114 | 54 | 255 | 109 | 39 | 199 | 67 | 216 | 141 | 41 | 218 | 210 | 165 | 62 | 140 | 22 |
| 138 | 88 | 215 | 97 | 45 | 193 | 90 | 20 | 153 | 158 | 24 | 35 | 235 | 112 | 34 | 19 |
| 157 | 248 | 77 | 59 | 204 | 25 | 173 | 94 | 238 | 9 | 121 | 201 | 23 | 233 | 99 | 70 |
| 118 | 98 | 102 | 254 | 169 | 106 | 243 | 174 | 185 | 230 | 164 | 184 | 144 | 160 | 195 | 182 |
| 110 | 103 | 142 | 183 | 172 | 239 | 7 | 127 | 225 | 208 | 1 | 104 | 36 | 129 | 131 | 75 |
| 147 | 202 | 150 | 66 | 85 | 92 | 100 | 0 | 60 | 167 | 179 | 69 | 152 | 16 | 209 | 3 |
| 29 | 120 | 122 | 44 | 181 | 12 | 50 | 13 | 212 | 188 | 49 | 177 | 47 | 203 | 32 | 237 |
| 200 | 227 | 61 | 4 | 71 | 232 | 64 | 186 | 15 | 72 | 81 | 58 | 154 | 52 | 252 | 229 |
| 220 | 56 | 119 | 105 | 221 | 132 | 11 | 79 | 180 | 111 | 253 | 139 | 224 | 236 | 187 | 176 |
| 27 | 107 | 249 | 197 | 213 | 28 | 192 | 196 | 126 | 128 | 124 | 37 | 205 | 143 | 42 | 53 |
| 18 | 135 | 189 | 48 | 17 | 8 | 55 | 223 | 163 | 46 | 5 | 134 | 26 | 30 | 228 | 190 |
| 2 | 78 | 219 | 6 | 211 | 175 | 161 | 250 | 101 | 166 | 10 | 108 | 82 | 123 | 226 | 116 |

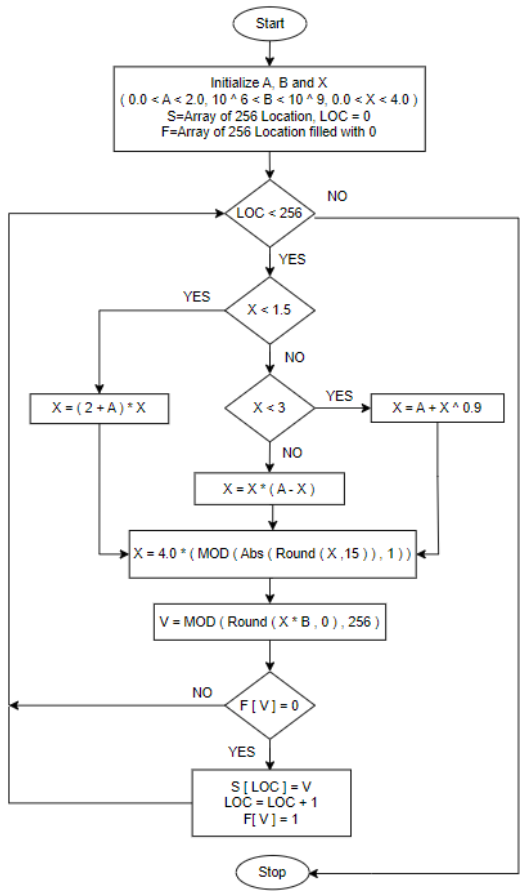

FIGURE 4: S-Box construction process.

this criterion, which was proposed by Tavares and Webster, any change in the input bits should cause the output bits to change separately. The proposed S-BIC-NL Box's values are presented in Table 6, with an average BIC-NL value of 103.6. The SAC and BIC-NL values of several S-Boxes are compared in Table 6, which highlights the high performance of the suggested S-Box in terms of both criteria.

TABLE 8: BIC-NL values of proposed S-Box.

| 0 | 106 | 102 | 100 | 104 | 104 | 104 | 102 |
|---|---|---|---|---|---|---|---|
| 106 | 0 | 100 | 104 | 104 | 106 | 104 | 100 |
| 102 | 100 | 0 | 104 | 104 | 108 | 106 | 100 |
| 100 | 104 | 104 | 0 | 100 | 104 | 104 | 102 |
| 104 | 104 | 104 | 100 | 0 | 106 | 108 | 106 |
| 104 | 106 | 108 | 104 | 106 | 0 | 102 | 106 |
| 104 | 104 | 106 | 104 | 108 | 102 | 0 | 102 |
| 102 | 100 | 100 | 102 | 106 | 106 | 102 | 0 |

### D. DIFFERENTIAL APPROXIMATION PROBABILITY (DP)

Differential cryptanalysis is a form of assault on symmetric-key block ciphers like the Data Encryption Standard that was developed by Biham and Shamir in 1990. (DES). It also holds true for other ciphers that employ permutations and substitutions. An S-resistance Box for differential cryptanalysis is evaluated using differential uniformity (DU) and differential probability (DP) values. The equation can be used to compute the DU of a certain S-Box (8).

$$DU = MAX_{\triangle c \neq 0, \triangle y}[\#\{c \in N \mid S(c) \oplus S(c \oplus \triangle c) = \triangle y\}] \quad (8)$$

Where "N" stands for all potential inputs. The proposed S-Box displays a low DU value of 0.039, which highlights its strong resilience to attacks utilizing differential cryptanalysis. The DU values for the provided S-Box are listed in Table 7. A comparison of the DP values for various S-Boxes is shown in Table 8.

### E. BIT INDEPENDENCE CRITERION (BIC)

Another crucial factor for assessing the effectiveness of S-Boxes is the Bit Independence Criterion (BIC). According to

TABLE 9: Differential uniformity values of projected S-Box.

| 6 | 8 | 6 | 6 | 6 | 8 | 6 | 8 | 6 | 8 | 4 | 8 | 6 | 6 | 8 |
|---|---|---|---|---|---|---|---|---|---|---|---|---|---|---|
| 8 | 6 | 6 | 6 | 6 | 6 | 6 | 8 | 6 | 8 | 6 | 8 | 8 | 8 | 6 | 8 |
| 4 | 8 | 6 | 8 | 8 | 6 | 6 | 8 | 6 | 8 | 8 | 6 | 6 | 6 | 6 | 8 |
| 6 | 8 | 6 | 8 | 10 | 6 | 6 | 8 | 6 | 6 | 6 | 8 | 8 | 8 | 6 | 8 |
| 6 | 8 | 8 | 8 | 6 | 8 | 8 | 8 | 8 | 6 | 6 | 6 | 6 | 6 | 6 |
| 6 | 10 | 8 | 6 | 6 | 8 | 6 | 8 | 6 | 8 | 8 | 8 | 6 | 8 | 6 | 6 |
| 6 | 8 | 6 | 6 | 6 | 6 | 6 | 6 | 6 | 8 | 8 | 8 | 8 | 6 | 6 |
| 6 | 8 | 6 | 6 | 8 | 6 | 6 | 8 | 8 | 6 | 8 | 6 | 6 | 6 | 8 |
| 8 | 6 | 6 | 6 | 8 | 6 | 6 | 8 | 6 | 10 | 10 | 6 | 6 | 6 | 8 | 6 |
| 8 | 6 | 6 | 6 | 6 | 6 | 6 | 8 | 6 | 6 | 8 | 4 | 8 | 10 | 6 |
| 8 | 8 | 6 | 6 | 6 | 8 | 8 | 8 | 8 | 6 | 6 | 6 | 8 | 6 | 6 |
| 6 | 8 | 10 | 8 | 6 | 6 | 8 | 6 | 6 | 8 | 8 | 8 | 8 | 8 | 8 | 8 |
| 6 | 6 | 8 | 8 | 6 | 6 | 6 | 8 | 6 | 8 | 6 | 6 | 8 | 8 | 8 | 6 |
| 6 | 8 | 4 | 6 | 6 | 8 | 6 | 8 | 8 | 8 | 6 | 6 | 6 | 6 | 6 |
| 6 | 6 | 6 | 8 | 6 | 6 | 8 | 6 | 4 | 6 | 8 | 6 | 6 | 6 | 6 |
| 8 | 6 | 6 | 8 | 6 | 6 | 8 | 8 | 10 | 6 | 8 | 10 | 6 | 6 | 6 | 0 |

### V. EFFICIENCY ANALYSIS

On a Windows 8 machine with 4GB RAM and an Intel Core i7 processor running at 2.2 GHz, a simulation was run to assess the computational effectiveness of the suggested S-Box approach. The effectiveness of the initial and final S-Boxes was tested. To strengthen the cryptography of the initially formed S-Box, a novel, and heuristic approach is used to generate the final S-Box. To estimate the time needed to create the final S-Box with various parameter values, the time complexity of 100,000 different beginning S-Boxes was measured. Despite the great speed of contemporary CPUs, data security should not be compromised because it is a significant issue. Figure 5 shows how the novel heuristic approach against computational time results in an improvement in the nonlinearity of the first S-box.

### A. KEYSPACE

The suggested method for creating S-boxes is dynamic and key-dependent, enabling the creation of new S-boxes with various initial parameter values. Table 10 lists the parameters utilized in the method, their corresponding ranges, and key spaces. The overall key space of the proposed technique is approximately

$$6 * 10^{81} \sim 2^{272}$$

Which is an extremely large space for any attacker. As a result, the proposed technique is highly resistant to brute force attempts by potential attackers.

TABLE 10: KeySpace.

| Parameter | Parameter and Range | Key Space |
|---|---|---|
| X | 0 < X < 4.0 (15 decimal digits) | $4 * 10^{15}$ |
| A | 0 < A < 2.0 (15 decimal digits) | $2 * 10^{15}$ |
| B | 1000000 < B < 1000000000 | $10^3$ |
| C | 0 < C < 1000000000 | $10^9$ |
| D | 0 < D < 1000000000 | $10^9$ |
| E | 0 < A < 1.0 (15 decimal digits) | $10^{15}$ |
| F | 0 < A < 1.0 (15 decimal digits) | $10^{15}$ |

### VI. LIMITATIONS OF THE PROPOSED CHAOTIC MAP

The proposed technique employs a novel chaotic map for generating dynamic S-boxes, which can be used in creating new ciphers. However, it should be noted that this chaotic map has a limitation in terms of its static dimensionality, which is only one. Therefore, it is unclear how the impact of chaotic maps on performance would scale with more dimensions. Furthermore, the comparison of the proposed chaotic map has only been made with Logistic and Sine maps, and a more comprehensive comparison with other chaotic maps could potentially lead to further improvements in the results obtained.

### VII. CONCLUSION

In this study, a brand-new key-dependent substitution box is proposed. It was made using a dynamic chaotic map and a permutation process. Both of these new techniques are dynamic in nature and derive the values of their parameters from the encryption key. The collection of values can change even little and yet produce a new S-Box. Through comparison and investigation, it is demonstrated that the proposed chaotic map possesses a high level of chaotic complexity. Standard cryptographic standards are used to assess the designed S-Box's strength and to compare it to other S-Boxes based on chaotic maps. The comparison demonstrates how well-suited the designed S-Box is for cryptography applications.

. . .